\documentclass[aps,amsmath,twocolumn,amssymb,floatfix,showpacs,superscriptaddress,nofootinbib,longbibliography]{revtex4-1}
\usepackage{mathtools}
\usepackage{braket}
\usepackage[dvipsnames]{xcolor}
\usepackage{float}
\usepackage{physics}
\usepackage{subfigure}
\usepackage{graphics}
\usepackage{bm}
\usepackage{soul}
\usepackage{amsfonts,amsmath,amsthm,amssymb}
\usepackage{eucal}
\usepackage{MnSymbol}
\usepackage{braket}
\usepackage{float}
\usepackage{tikz}
\usepackage{blkarray}
\usepackage{pgffor}
\usepackage{float}
\usepackage{multirow}
\usepackage{booktabs}
\usepackage{silence}
\usepackage{makecell}
\usepackage[colorlinks=true,linktoc=page,linkcolor=BrickRed,citecolor=magenta,urlcolor=purple]{hyperref}

\mathchardef\mhyphen="2D 

\newcommand\bea{\begin{eqnarray}}
\newcommand\eea{\end{eqnarray}}
\newcommand\beq{\begin{equation}}  
\newcommand\eeq{\end{equation}}

\usepackage[normalem]{ulem}
\definecolor{lime}{HTML}{A6CE39}
\usepackage{sidecap,tikz}
\DeclareRobustCommand{\orcidicon}{\hspace{-1.0mm}
	\begin{tikzpicture}
		\draw[lime, fill=lime] (0.0,0.0) 
		circle [radius=0.15] 
		node[white] {{\fontfamily{qag}\selectfont \tiny \,ID}};
		\draw[white, fill=white] (-0.0525,0.095) 
		circle [radius=0.007];
	\end{tikzpicture}
	\hspace{-3.0mm}
}
\foreach \x in {A, ..., Z}{\expandafter\xdef\csname orcid\x\endcsname{\noexpand\href{https://orcid.org/\csname orcidauthor\x\endcsname}
		{\noexpand\orcidicon}}
}

\AtBeginDocument{%
	\newwrite\bibnotes
	\def\bibnotesext{Notes.bib}
	\immediate\openout\bibnotes=\jobname\bibnotesext
	\immediate\write\bibnotes{@CONTROL{REVTEX41Control}}
	\immediate\write\bibnotes{@CONTROL{%
			apsrev41Control,author="08",editor="1",pages="1",title="1",year="1"}}
	\if@filesw
	\immediate\write\@auxout{\string\citation{apsrev41Control}}%
	\fi
}%
\begin{document}

\title{Decoherence of $q-$Deformed Photon Added Coherent
State}
\author{Amit Das\orcidA{}}
\email{amit.rs@presiuniv.ac.in}
\affiliation{Department of Physics, Presidency University, 86/1 College Street, Kolkata, 700073, West Bengal, India}

\author{Sobhan Sounda}
\email{sobhan.physics@presiuniv.ac.in}
\affiliation{Department of Physics, Presidency University, 86/1 College Street, Kolkata, 700073, West Bengal, India}

\vspace{0.8cm}
\begin{abstract}
In this study, we explore the behavior of photon added coherent states in a deformed harmonic oscillator subjected to dissipative decoherence. We use $q-$deformation as our nonlinear function to model our system. By adjusting the deformation parameter, we show that $q-$deformed photon added coherent state (DPACS) exhibit greater nonclassicality and resilience to decoherence compared to those of a standard harmonic oscillator. Additionally, we investigate the nonclassical properties and entanglement of DPACS under decoherence induced by interaction with a dissipative photon-loss environment.
\end{abstract}

\maketitle

\section{Introduction}
\label{intro}
Coherent states corresponding to different Lie algebras have been extensively studied in many branches of physics. A coherent state in context of quantum optics follows Heisenberg–Weyl algebra. Coherent states constructed from $su(1,1)$, $su(2)$, $su_q(2)$ algebra also found useful in many applications. It is worth to mention that the generalized coherent states can be constructed from three definitions, (i) $\ket{\alpha}$ is the right eigen state of the boson anhilation operator ($\hat{a}$) ie, $\hat{a}\ket{\alpha}=\alpha\ket{\alpha}$ (ii) applying unitary displacement operator $\hat{D}(\alpha)=exp{(\alpha\hat{a}^\dag -\alpha^* \hat{a})} $ to the vacuum state $\ket{0}$ and (iii) the states are the minimum uncertainty state. The states satisfying all three of these conditions are called intelligent states, \cite{Aragone1974,Trifonov1994}. Optical coherent state $\ket{\alpha}$ is one of these unique intelligent state that closely mimics the behavior of a classical electromagnetic field \cite{Glauber2,Sudarshan2}.  
Over the last forty years, numerous studies have focused on coherent states and their use across various areas of physics, explored from both theoretical and experimental perspectives. The construction and application of coherent states led to various mathematical formulation of generalization of coherent states  \cite{Gilmore1972,Perelomov1972,Perelomov1986,Gerry2005a,Markham2003}.\\
Agarwal and Tara introduced the photon added coherent state \cite{Agarwal1991} as the excitation of the coherent state. The single-photon-added coherent state has been experimentally generated and
characterized \cite{Zavatta2004} using quantum homodyne tomography. This state establishes a connection between the ideal coherent state, which characterizes the classical nature of the radiation field, and the Fock state which is highly nonclassical. Hence various photon added generalized coherent states \cite{Naderi2004,Safaeian2011,Faghihi2020,Deyy2016} have been studied extensively.\\
With the advancement of experimental techniques, there has been significant interest in observing nonclassical phenomena of light, such as photon antibunching \cite{Short1983} , Wigner negativity \cite{Lvovsky2000}, squeezing \cite{Slusher1985,Gerry2005b} and entanglement \cite{Altepeter2005} . Thus the generation and realization of nonclassical states of light have attracted considerable attention due to their applications in quantum communication \cite{Zugenmaier2018} , quantum cryptography \cite{Usenko2018,Wang2019}, and quantum information processing \cite{Braunstein2005}.\\
Generalizations in quantum optics and its mathematical structure for nonlinear systems can be done several ways. Some of the well-known approaches  including work by R. L. de Matos Filho and W. Vogel \cite{MatosFilho1996} and $f-$oscillator approach by Man’ko et al \cite{Manko1997} and q-deformation \cite{Biedenharn1989b,Macfarlane1989}. A common objective of these approaches is to construct quantum optical states associated with potentials beyond the harmonic oscillator. Importantly, generalized nonclassical states have been found to possess a greater degree of nonclassicality compared to standard nonclassical states \cite{Dey2015,Dey2016}, and this feature can be used in many quantum information application \cite{Kim2002,DeyHussin2015}.\\
Quantum states allow coherent superpositions, unlike classical states, which are described by statistical mixtures. However, this nonclassical feature caused by the quantum superposition is difficult to preserve in the presence of interactions with the external environment. Such interactions lead to a loss of coherence in the state,  known as decoherence \cite{Gardiner1991}. The process of Decoherence not only depends on the state that is prepared \cite{Belfakir2020,Berrada2019} also can be useful in studying about the interaction medium \cite{Nandi2026}. This decoherence effect has been studied extensively, both theoretically and experimentally, in various physical systems \cite{Myatt2000,Kis2001,Mancini2001}. \\ Thus motivated by these discussions above our aim in this work is to study the deformed photon added coherent state (DPACS) under decoherence. We first construct the DPACS using the the $f-$oscillator approach. Due to the equivalence among different approaches for constructing generalized coherent states \cite{Roknizadeh2004}, we adopt the $f-$oscillator method for convenience. We choose $q-$deformation as our nonlinear model function. We choose the interaction  environment as the simplest photon loss channel at zero temperature. Then we study the evolution of various nonclassical measures to see the effect of decoherence on our state DPACS.
\\
The organization of the paper is as follows: In Sec.~\ref{II} we recall the $f-$oscillator algebra and construct the $q-$DPACS and discuss the decoherence due to the dissipative photon loss channel. In Sec.~\ref{III} we calculate and discuss various nonclassical indicator like Mandel's $\mathcal{Q}$ Parameter, Wigner Function etc and their evolution due to decoherence. Finally in Sec.~\ref{IV} we mention the conclusion of our work.
\section{Deformed Photon added coherent state}
\label{II}
We adopt the nonlinear coherent states framework or the \textit{f-}oscillator approach, in which the conventional bosonic annihilation $\hat{a}$ and creation $\hat{a}^\dag$ operators are modified through a nonlinearity function. The \textit{f-}oscillator operator are defined by
 \begin{equation}
     \hat{A}=\hat{a}f(\hat{n})=f(\hat{n}+1)\hat{a}\quad,\quad \hat{A}^\dag=f(\hat{n})\hat{a}^\dag=\hat{a}^\dag f(\hat{n}+1)
 \end{equation}
 here $\hat{A}$ and $\hat{A}^\dag$ are the \textit{f-}deformed anhilation and creation operator and $\hat{n}=\hat{a}^\dag\hat{a}$ is the number operator. These operator satisfy the following nonlinear algebra
 \begin{align}
     [\hat{A},\hat{A}^\dag]=(\hat{n}+1)f^2(\hat{n}+1)-\hat{n}f^2(\hat{n})\\ \nonumber
     [\hat{N},\hat{A}]=-\hat{A} \quad , \quad [\hat{N},\hat{A}^\dag]=\hat{A}^\dag
 \end{align}
 Clearly for $f(\hat{n})=1$ the nonlinear algebra reduces to the usual Heisenberg algebra.\\
 It is worth mentioning that the action of the \textit{f-}deformed creation operator on the Fock basis
 \begin{equation}\label{eq2}
     \hat{A}^{\dag m} \ket{n}=D_{m,n}\ket{n+m} \text{~~where,~~}  D_{m,n}=\prod_{i=n+1}^{n+m}f(i)\cdot \sqrt{i}
 \end{equation}
 Similar to the usual coherent states the nonlinear coherent states are defined as the right eigen-state of the \textit{f-}deformed annihilation operator $\hat{A}$
 \begin{equation}\label{eq1}
     \hat{A}\ket{\alpha,f}=\alpha\ket{\alpha,f}, \qquad \alpha \in \mathbb{C}
 \end{equation}
 In Fock state representation this Eq.~\ref{eq1} leads to 
 \begin{equation}
     \ket{\alpha,f}=\mathcal{N}_o \sum_{n=0}^{\infty}\alpha^n\cdot g_n\ket{n}
 \end{equation}
 here $\mathcal{N}_o$ is the normalization constant.
 \begin{align}
     g_n&=\frac{1}{\sqrt{n!}\cdot f(n)!}\qquad \text{where,} \quad f(n)!=\prod_{i=0}^{n}f(i)  \nonumber\\
     \mathcal{N}_o&=\Bigl[\sum_{n=0}^{\infty}g_n^2 \cdot \abs{\alpha}^{2n}\Bigr]^{-\frac{1}{2}}
 \end{align}

 Now in our study we consider the deformed photon added coherent state (DPACS) which is defined by 
 \begin{equation}
  \ket{\psi}=\mathcal{N}_+ \hat{A}^{\dag m} \ket{\alpha,f}
 \end{equation}
  In Fock basis the expansion of the DPACS using Eq.~\ref{eq2}
 \begin{align} \label{psi}
     \ket{\psi}&=\mathcal{N}_+ \mathcal{N}_o \sum_{n=0}^{\infty}\alpha^n\cdot g_n \cdot D_{m,n}\ket{n+m} \\
     \mathcal{N}_+&=\Bigl[\mathcal{N}_o^2\sum_{n=0}^{\infty}\abs{\alpha}^{2n}\cdot g_n^2\cdot D_{m,n}^2\Bigr]^{-\frac{1}{2}} \nonumber
 \end{align}
\begin{figure}[ht!]
\centering
(a)\includegraphics[width=0.8\linewidth]{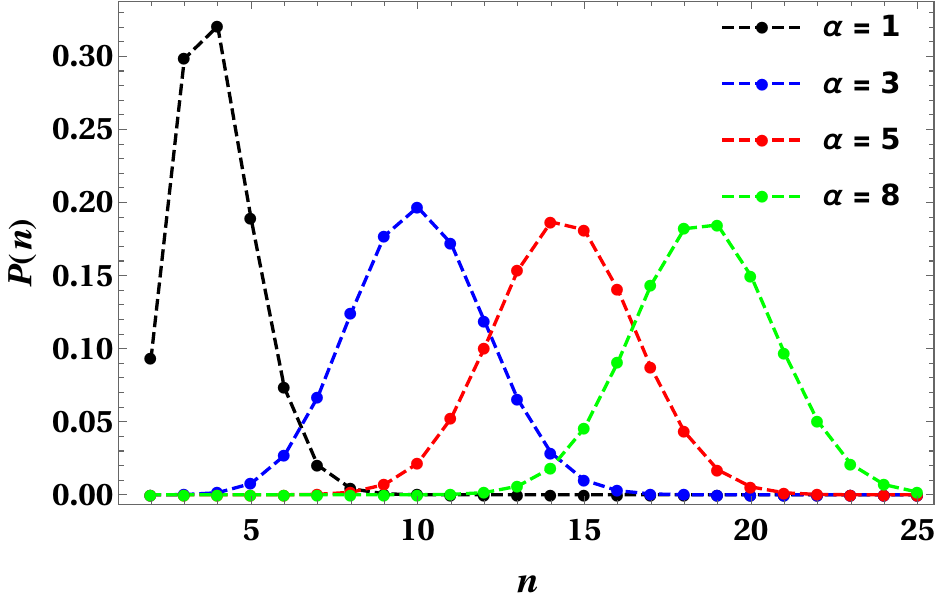}
(b)\includegraphics[width=0.8\linewidth]{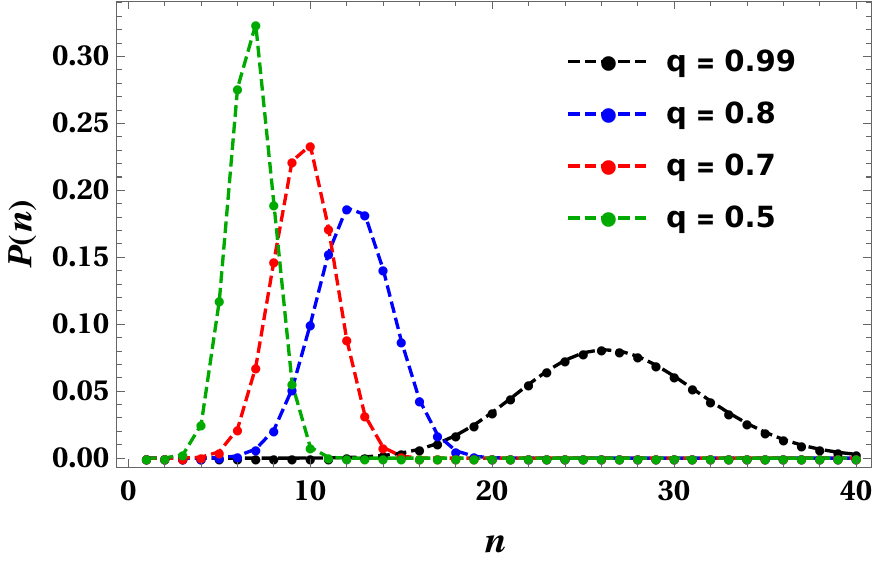}
~~(c)\includegraphics[width=0.8\linewidth]{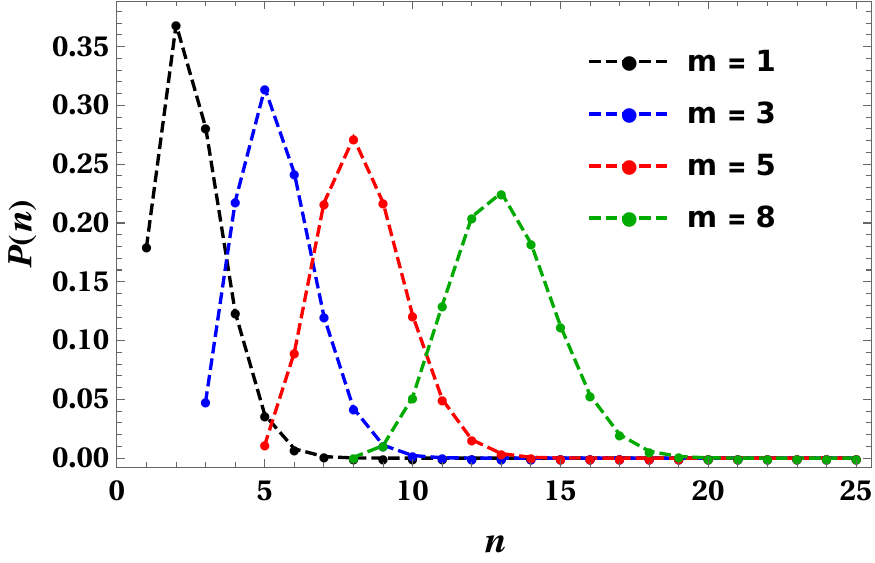}

 \caption{Photon number distribution(PND) $P_n(0)$ of DPACS , considering $q-$deformation oscillator in absence of the dissipative photon loss environment ,(a) PND for $q=0.8$ and $m=2$ for different values of $\alpha$ (b) PND for $\alpha=5$ and $m=1$ for different values of $q$ (c) PND for $q=0.8$ and $\alpha=1$ for different values of $m$}
    \label{fig=1}
\end{figure}
 Now for the sake of simplicity we use the notation in a simple form
\begin{equation}\label{rho1}
     \ket{\psi}=\sum_{n=0}^{\infty}C_n \cdot\ket{n}
 \end{equation}
 where \begin{equation}
C_n =\begin{cases}(\mathcal{N}_+ \mathcal{N}_o)\times\alpha^{\,n-m}\, d_{n-m}\, D_{m,n-m}, & n\ge m, \\[6pt]0, & n < m .
\end{cases}
\end{equation}
\begin{figure*}[ht!]
    \centering
    (a)\includegraphics[width=0.4\linewidth]{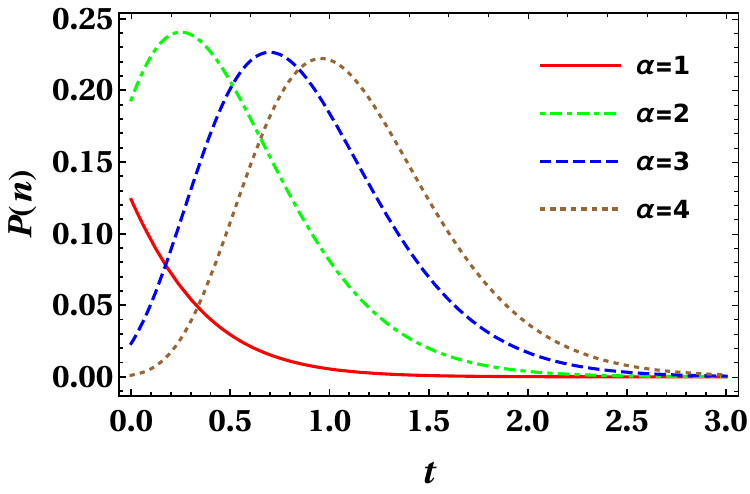}
    (b)\includegraphics[width=0.4\linewidth]{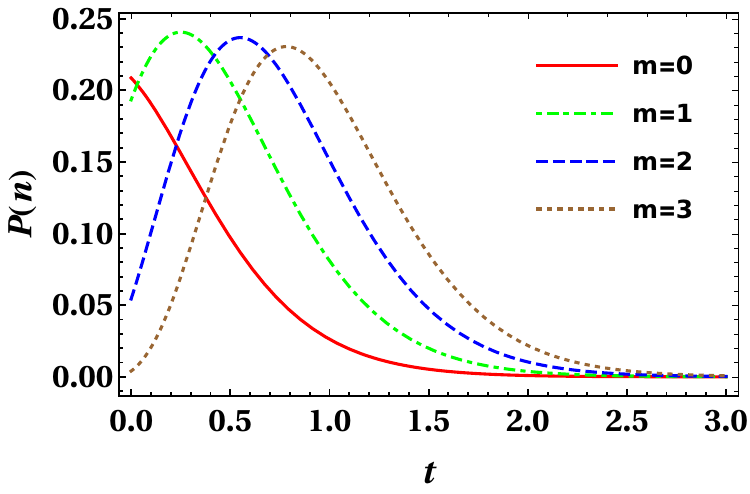}
    (c)\includegraphics[width=0.4\linewidth]{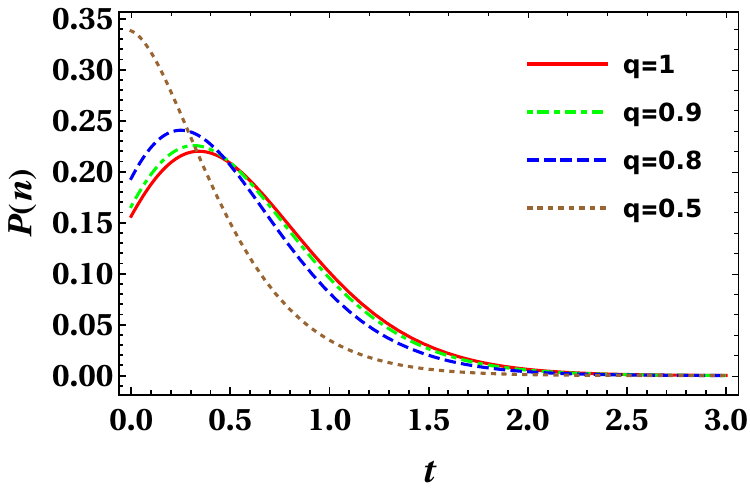}
    (d)\includegraphics[width=0.4\linewidth]{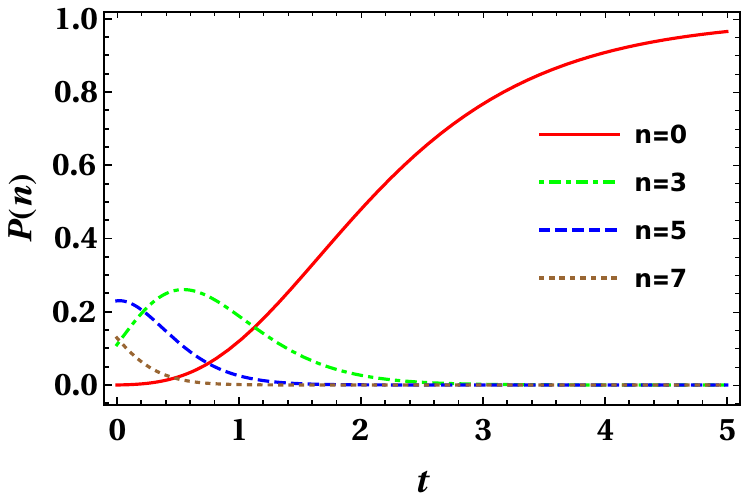}
    (e)\includegraphics[width=0.4\linewidth]{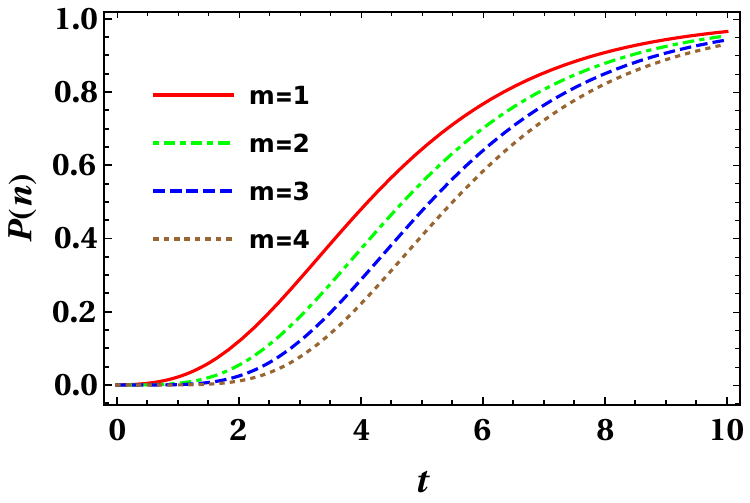}
    (f)\includegraphics[width=0.4\linewidth]{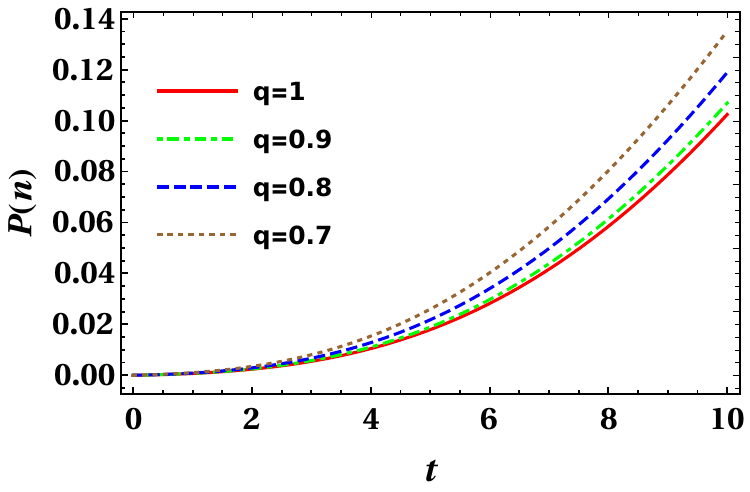}
    \caption{Time evolution of the photon distribution function $P_n(t)$ for a range of deformation parameters of DPACS, considering $q-$deformation oscillator (a) $n=4,m=1,q=0.8,\gamma=1$ for different values of $\alpha$ (b) $n=4,\alpha=2,q=0.8,\gamma=1$ for different values of $m$ (c) $n=4,m=1,\alpha=2,\gamma=1$ for different values of $q$ (d) $m=1,\alpha=2,q=0.8,\gamma=1$ for different values of $n$  (e) $n=0,q=0.8,\alpha=2,\gamma=0.5$ for different values of $m$  (f) $n=0,m=1,\alpha=2,\gamma=0.1$ for different values of $q$ }
    \label{fig:PN(T)}
\end{figure*}
The procedure we have adopted is general except the deformation function $f(\hat{n})$. The deformation function $f(\hat{n})$ arose from many solvable quantum systems or states constructed from group theoratical approach. For example Hydrogen like spectrum \cite{Honarasa2009}, center of mass motion of trapped ions \cite{MatosFilho1996}, Pöschl–Teller potential \cite{Shreecharan2004},Barut–Girardello coherent states of $SU(1,1)$ group \cite{Barut1971}, Gilmore–Perelomov coherent states of $SU(1,1)$ group \cite{Perelomov1972} could be discussed in this context. In our case we have studied the $q-$deformed case. In the present work, the deformation function $f(n)$ is taken to follow from the algebra satisfied by the canonical variables associated with the $q-$deformed oscillator \cite{Manko1997,Manko1993IJMPA,Biedenharn1989}.
\begin{equation}
    f(n)=\sqrt{\frac{1}{n}\frac{q^n-q^{-n}}{q-q^{-1}}}
\end{equation}
With this deformation function our deformed oscillator algebra comes down to the nice structure \begin{equation}
    \hat{A}^{\dag}\ket{n}=\sqrt{[n+1]_q}\ket{n+1} \text{~~and~~}\hat{A}\ket{n}=\sqrt{[n]_q}\ket{n-1}
\end{equation}  where 
\begin{equation}
    [n]_q=\frac{q^n-q^{-n}}{q-q^{-1}}=nf^2(n)
\end{equation}
 Following to this we want to see the behavior of this DPACS due to photon loss in a dissipative environment. The decoherence effect due to the interaction with the dissipative media can be described by the master equation of the Lindblad form (setting bath temperature to zero)
 \begin{equation}
     \dv{\hat{\rho}(t)}{t}=\gamma \hat{a}\hat{\rho}(t)\hat{a}^\dag-\frac{\gamma}{2}\bigl\{\hat{a}^\dag \hat{a},\hat{\rho}(t)\bigr\}
 \end{equation}
 Thus the decoherence induced by photon loss on the state $\hat{\rho}(0)=\op{\psi}{\psi}$ can be characterized as follows
 \begin{equation}
     \hat{\rho}(t)=\sum_{i=0}^{\infty}\hat{S}_i(t)\hat{\rho}(0)\hat{S}_i^\dag(t) \qquad 
     \label{damp1}
 \end{equation}
 $\hat{\rho}(t)$ is the reduced density matrix of the system. The Kraus operator $\hat{S}_i(t)$ is represented in fock basis as
 \begin{equation}
     S_i(t)=\sum_{n=i}^{\infty}\sqrt{\mqty(n \\ i)}[\eta(t)]^{\frac{(n-i)}{2}}[1-\eta(t)]^{\frac{i}{2}}\op{n-i}{n} 
     \label{kir1}
 \end{equation}
 here $\eta(t)=e^{-\gamma t}$ and $\gamma$ is the decay rate of the bath.

\section{DECOHERENCE}
\label{III}
We now analyse various nonclassical properties and their dynamics under dissipative decoherence of the deformed PACS. 

\subsection{Photon Number Distribution}\label{PNDD}

The probability $P_n$ of having $n$  photons in the DPACS state is calculated by the formula 
\begin{equation}\label{PND_t}
    P_n(t)=Tr[\hat{\rho}(t)\op{n}{n}]
\end{equation}
So in absence of the interaction of the bath the $P_n(0)$ is 
\begin{equation}
    P_n(0)=\abs{C_n}^2
\end{equation}
The results for the photon number distribution 
$P_n(0)$ in the absence of the bath, for various parameters, are shown in Fig.~\ref{fig=1}. We find that $P_n(0)$ depends on the deformation parameter q through the relation $P_n(0) \propto \frac{[n]_q !}{[n-m]_q^2!}$.
From Fig.~\ref{fig=1}(b), it is evident that decreasing the parameter q leads to an overall suppression of the probabilities corresponding to higher Fock states. At the same time, the distribution becomes more localized, with the peak shifting toward lower photon numbers and increasing in magnitude. This indicates that the photon number distribution becomes more concentrated in the few-photon region as the deformation increases. A similar trend is observed in Fig.~\ref{fig=1}(c) with variation in the photon-addition parameter m. Increasing m shifts the distribution toward higher photon numbers while simultaneously modifying the shape and sharpness of the peak, reflecting the combined influence of photon addition and deformation on the structure of the state.

Inserting Eq.~\ref{psi} in Eq.~\ref{PND_t} we obtain the expression for the photon number distribution as

\begin{equation}\label{pnd2}
   P_n(t)= \sum_{k=0}^{\infty} \mqty(n+k \\k)[\eta(t)]^n [1-\eta(t)]^k \abs{C_{(n+k)}}^2
\end{equation}
\begin{figure}[ht!]
  \centering
   (a) \includegraphics[width=0.8\linewidth]{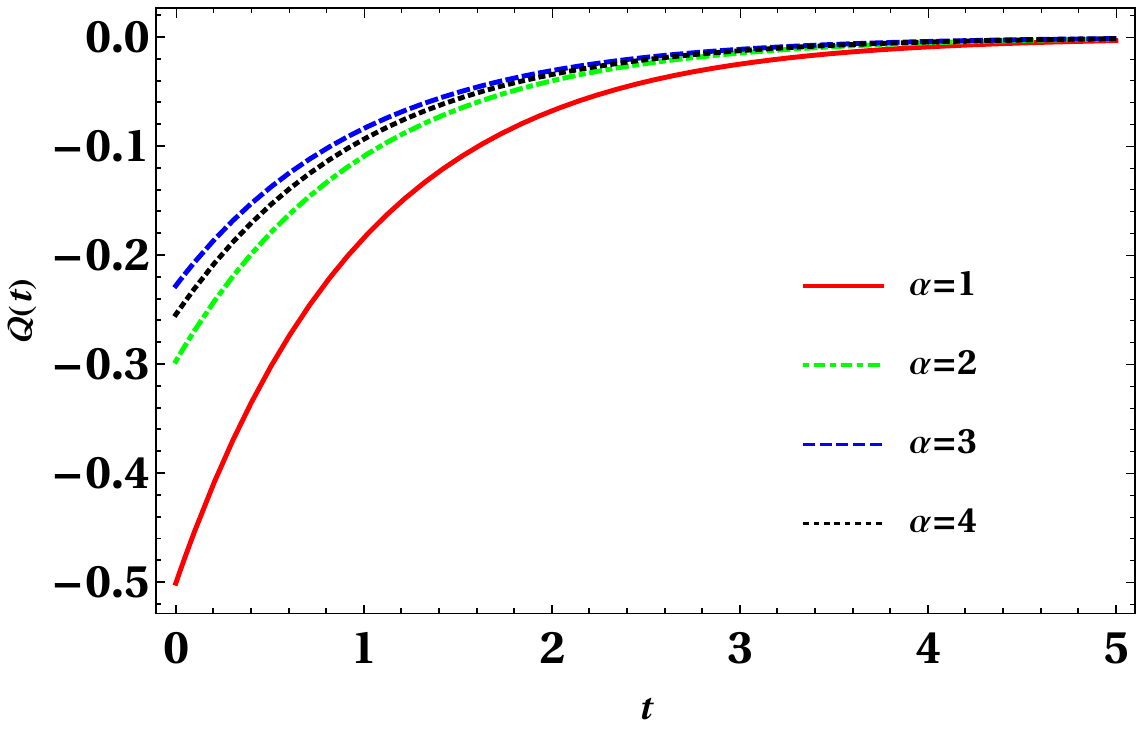}\\
    (b)\includegraphics[width=0.8\linewidth]{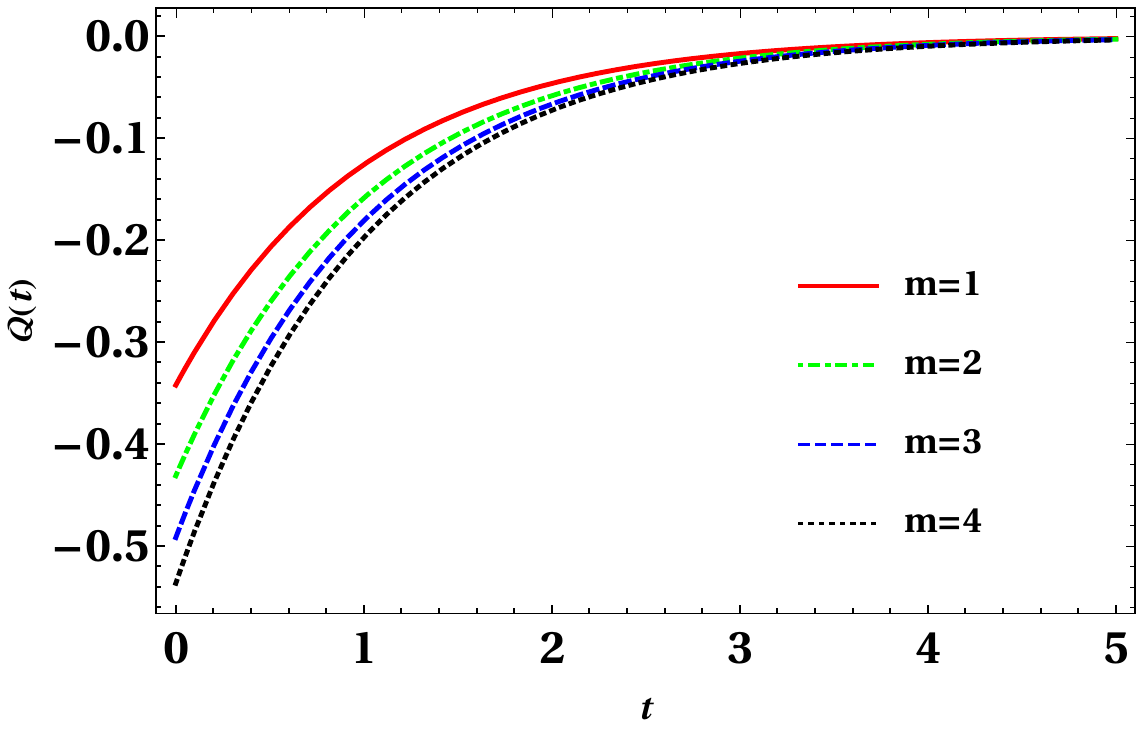}\\
    (c)\includegraphics[width=0.8\linewidth]{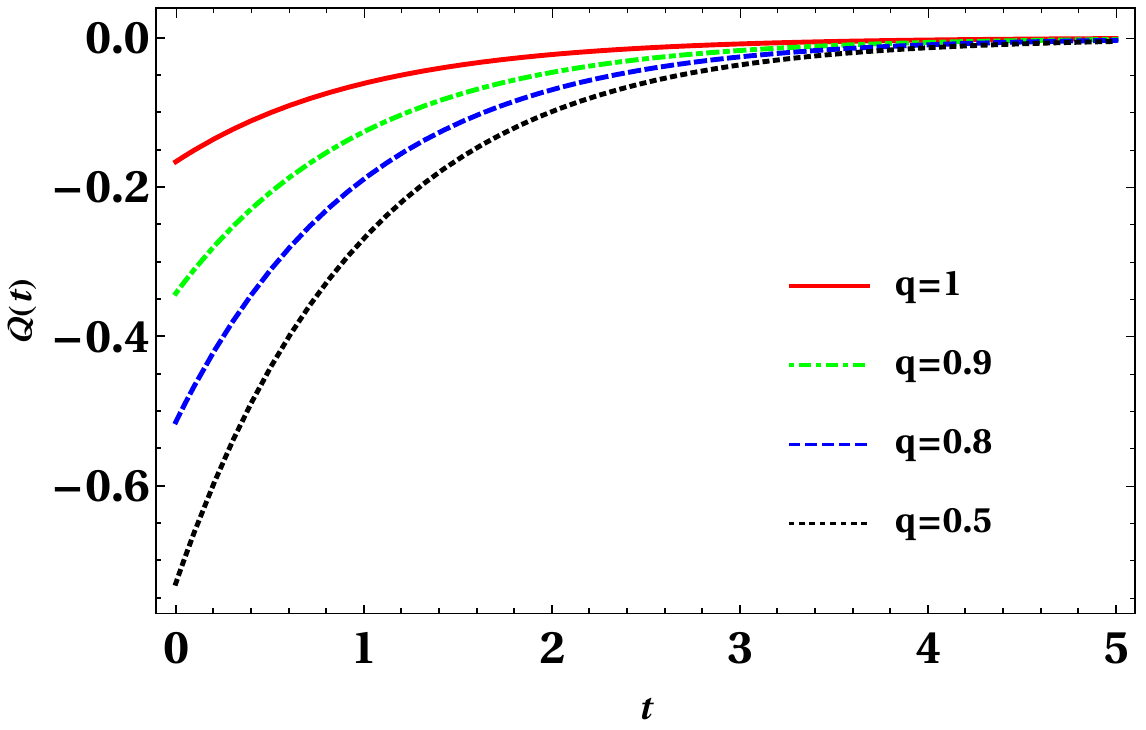}\\
   (d) \includegraphics[width=0.8\linewidth]{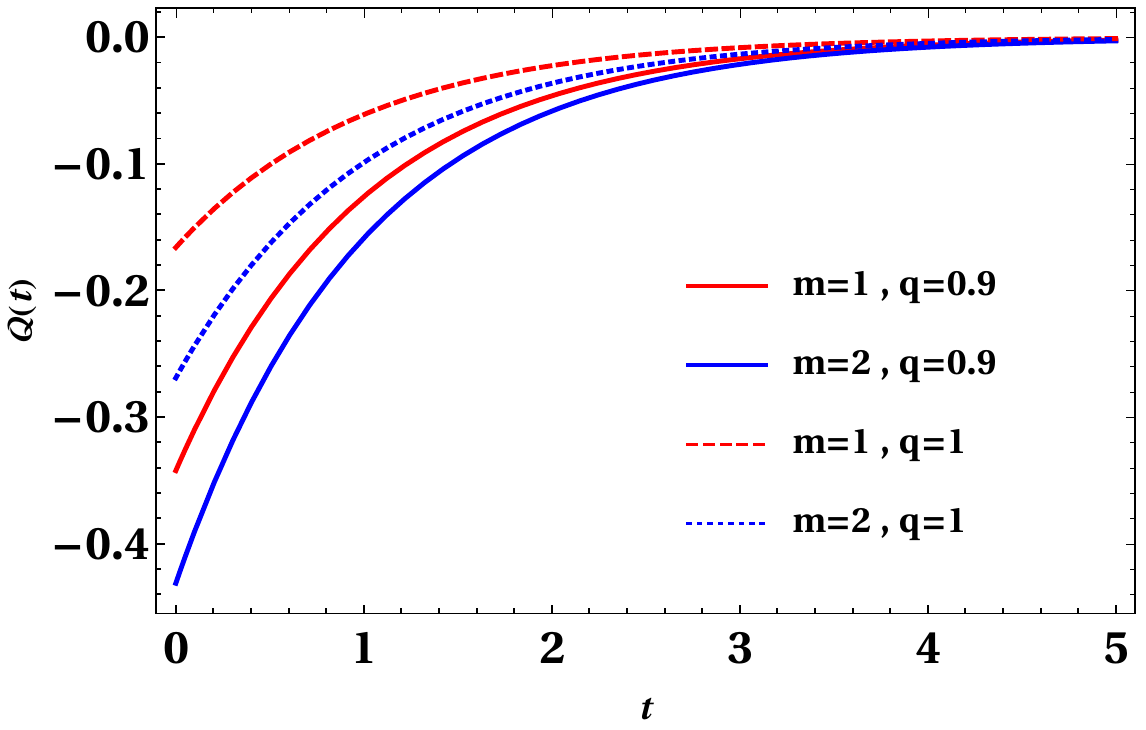}\\
    \caption{Time evolution (for $\gamma=1$) of the Mandel's $\mathcal{Q}$ Parameter for a range of deformation parameters, considering $q-$deformation oscillator (a) $m=1,q=0.95$ for different values of $\alpha$ (b) $\alpha=3,q=0.9$ for different values of $m$ (c) $m=1,\alpha=3$ for different values of $q$ (d) $\alpha=3$ for different values of $q$ and $m$ }
    \label{qpar}
\end{figure}

In Fig.~\ref{fig:PN(T)}(a),(b),(c) we can see the effect of parameters $\alpha$,$m$,$q$ respectively for the Fock basis number $n=4$ (we choose $\gamma=1$ for better visualization of the effect of decoherence ). From Fig.~\ref{fig:PN(T)}(c) we can see that more the deformation $q$ more the probability initially (at $t=0$), finally asymptotes to zero more quickly. Interesting to note from Fig.~\ref{fig:PN(T)}(d) under decoherence probability of all the Fock basis tends to zero except the vacuum state. The vacuum state $\ket{0}$ tends to 1 for large $t$. Thus we conclude under decoherence our state DPACS tends to the pure state that is the vacuum state , which is classical. Hence effect of parameters due to decoherence can be seen by looking the evolution of $P_{(n=0)}(t)$. So $P_{(n=0)}(t)$ seen from Fig.~\ref{fig:PN(T)}(e),(f) shows that increasing $m$ shows slower decoherence contrary to variation of $q$ which shows faster decoherence as our state DPACS is more deformed (ie, $q$ away from 1).

\subsection{Mandel's Q Parameter}\label{mandel_q}
Mandel’s $\mathcal{Q}$ parameter quantifies the sub-Poissonian nature of a quantum state, serving as an indicator of its nonclassical properties. Of particular interest is how this $\mathcal{Q}$ parameter evolves under the influence of decoherence. It is defined as follows
\begin{equation}
    \mathcal{Q}(t)=\frac{(\Delta \hat{n}(t))^2}{\expval{\hat{n}(t)}}-1= \frac{ \expval{\hat{n}^2(t)} - \expval{\hat{n}(t)}^2}{\expval{\hat{n}(t)}}-1
\end{equation}
where $\expval{\hat{n}(t)}$ is average value of the photon number, and $\expval{\hat{n}^2(t)}$ is the variance of photon number. The Mandel parameter  characterizes the photon-number statistics, taking the value $\mathcal{Q}=0$ for Poissonian distributions(most classical like quantum state), becoming positive $\mathcal{Q}>0$ for super-Poissonian statistics, and negative $\mathcal{Q}$ in the sub-Poissonian regime. The expectation values follows from, $\expval{\hat{n}^2(t)}=\sum_{n=0}^{\infty}n^2 P_n(t)$ and $\expval{\hat{n}(t)}=\sum_{n=0}^{\infty}n P_n(t)$. So the expression for the Mandel's $\mathcal{Q}$ as a function of time is
\begin{equation}
    \mathcal{Q}(t)=\frac{(\sum_{n=0}^{\infty}n^2 P_n(t))-(\sum_{n=0}^{\infty}n P_n(t))}{\sum_{n=0}^{\infty}n P_n(t)}-1
\end{equation}
Now using the expression fo $P_n(t)$ from Eq.~\ref{pnd2} we calculated the dynamics of the $\mathcal{Q}(t)$ parameter under the dissipative decoherence of DPACS which is shown in Fig.~\ref{qpar} for $\gamma=1$. From Fig.~\ref{qpar} for all the parameters the $\mathcal{Q}<0$, showing sub-Poissonian statistics and for large time  all states going towards $\mathcal{Q}=0$. Thus, the state undergoes a transition from a nonclassical to a classical regime under the influence of the dissipative photon-loss channel. This behavior is consistent with the photon number distribution analysis, which shows that the state gradually becomes dominated by the vacuum component. Since the vacuum is a classical Gaussian state, its increasing contribution drives the system toward classicality. Fig.~\ref{qpar}(b),(c) shows increase in nonclassicality as $m$ increases or deformation increases ( value of $q$ away from 1). As seen from Fig.~\ref{qpar}(d), for an identical number of added photons, the 
$q$-deformed state  displays more negativity in the Mandel parameter $\mathcal{Q}(t)$. This enhanced negative value signifies a stronger sub-Poissonian character compared to the undeformed case (ie, $q=1$), highlighting the role of $q$-deformation in amplifying nonclassical photon-number statistics.

\begin{figure*}
    \centering
    (a)\includegraphics[width=0.4\linewidth]{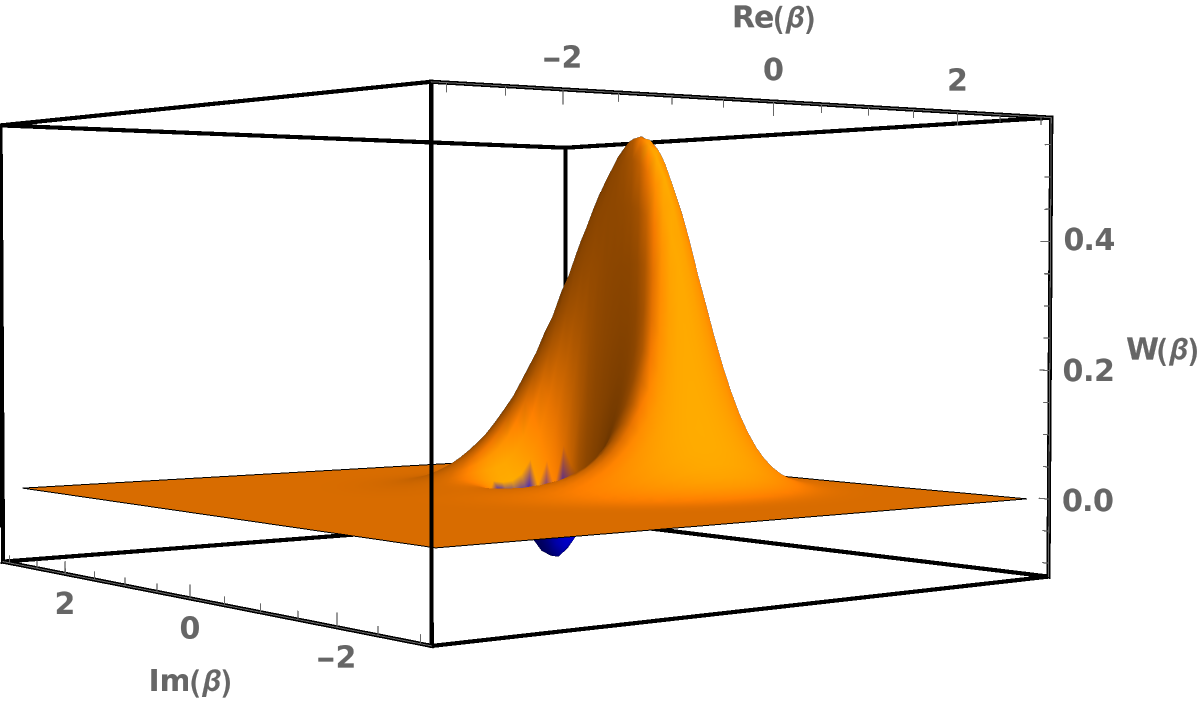}
    (b)\includegraphics[width=0.4\linewidth]{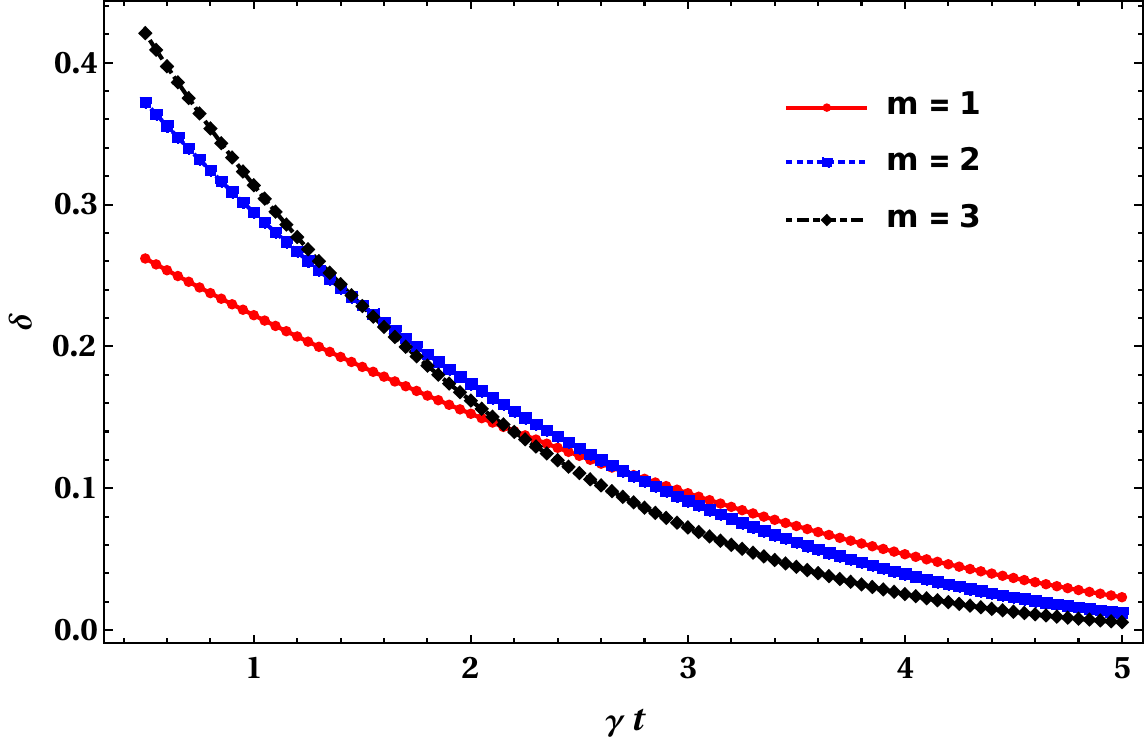}
    (c)\includegraphics[width=0.4\linewidth]{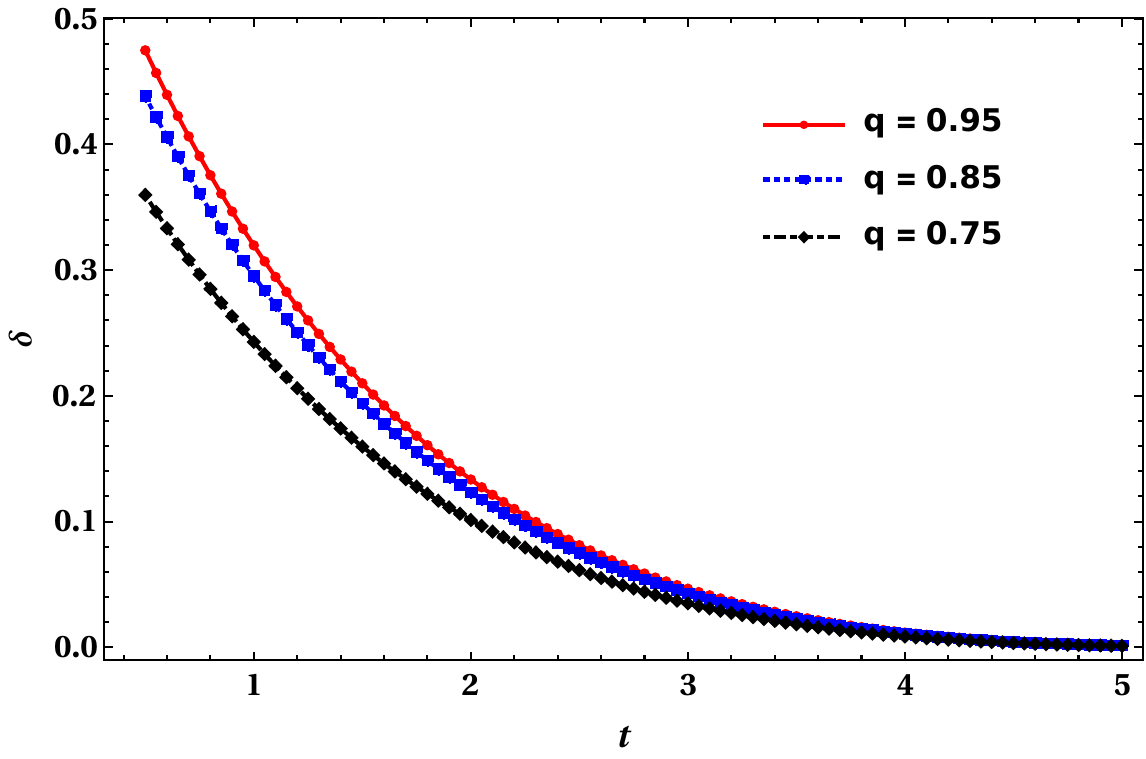}
   (d)\includegraphics[width=0.4\linewidth]{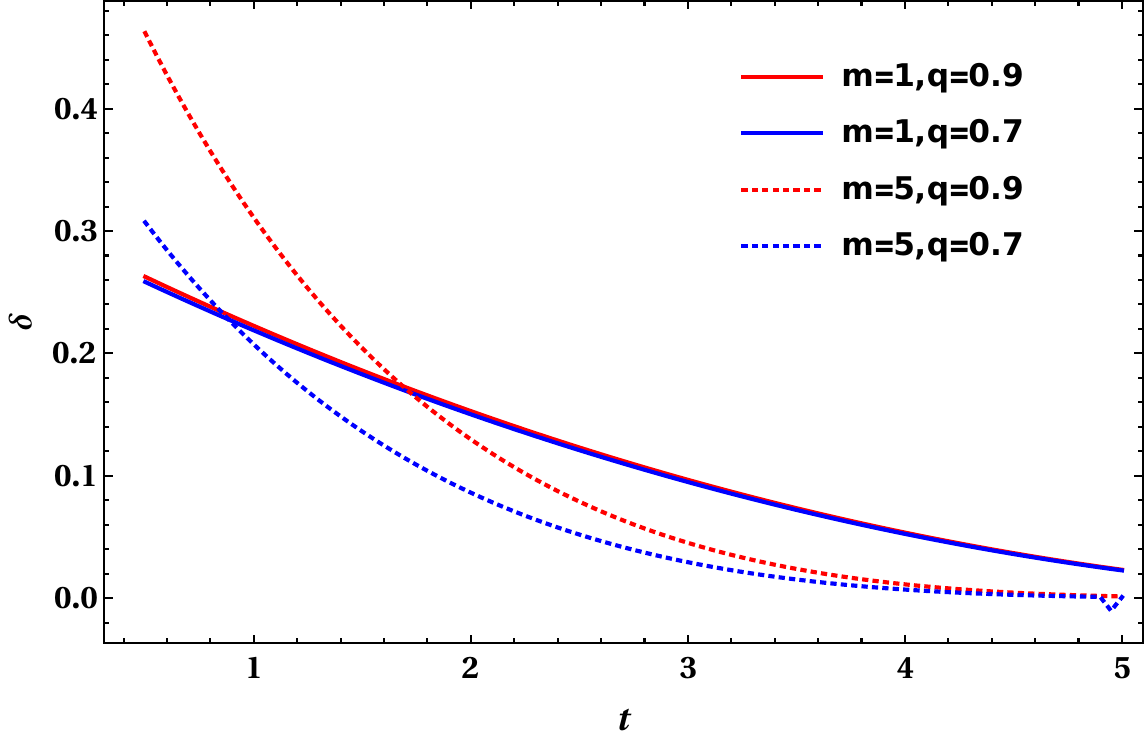}
   
    \caption{(a) Wigner function for $\alpha=1m=1,q=0.8,\gamma=0.5,t=0.5$. And time evolution of the negative volume of the Wigner function $\delta(t)$ for a range of deformation parameters, considering $q-$deformation oscillator (b) $\alpha=0.5,\gamma=0.1,q=0.85$ for different values of $m$ (c) $\alpha=0.5,\gamma=0.1,m=5$ for different values of $q$ (d) $\alpha=0.5,\gamma=0.1$ for different values of $q$ and $m$ }
    \label{fig:wigneg}
\end{figure*}

\subsection{Wigner function}
The Wigner function associated with a quantum state $\hat{\rho}(t)$ serves as a quasi-probability distribution in phase space. Its negative values are a clear signature of the state's nonclassical nature. In the phase space,  Wigner function for the state $\hat{\rho}(t)$ is defined as follows \cite{Royer1977,Wunsche2018}:
     
\begin{equation}
    W(\beta,t)=\frac{2}{\pi}Tr[\hat{\rho}\hat{D}(2\beta)\hat{\Pi}]
\end{equation}

where $\hat{\beta}=exp{(\beta \hat{a}^\dag-\beta^* \hat{a})}$ is the displacement operator, and $\hat{\Pi}=(-1)^{\hat{a}^\dag \hat{a}}$ is the parity operator. The Wigner function becomes
\begin{widetext}\label{wigfunc}
\begin{eqnarray}
W(\beta,t)=\frac{2}{\pi}\,e^{-2|\beta|^{2}}\sum_{n,m=0}^{\infty}
\rho_{n,m}\,(-1)^{\ell}\sqrt{\frac{\ell!}{(\ell+k)!}}\;\times
\begin{cases}(2\beta)^{k}\,L_{\ell}^{(k)}\!\left(4|\beta|^{2}\right),
& m \ge n, \\[6pt](-2\beta^{*})^{k}\,L_{\ell}^{(k)}\!\left(4|\beta|^{2}\right),& m < n .
\end{cases}
\label{wigfunc}
\end{eqnarray}
\end{widetext}

Where $l=min(m,n)$ and $k=\abs{n-m}$ and the $\rho_{n,m}$ is the matrix elements of the density matrix in Fock basis.
\begin{equation}
    \hat{\rho}(t)=\sum_{n,m=0}^{\infty}\rho_{n,m}(t)\op{n}{m}
\end{equation}
Using Eq.~\ref{rho1},~\ref{damp1},~\ref{kir1} we find the co-efficient of the density matrix under photon loss in the damping media $\rho_{n,m}(t)$ .
\begin{equation}
    \rho_{n,m}(t)=\sum_{i=0}^{\infty}C_{n+i}C^*_{m+i}\sqrt{\mqty(n+i \\ i)\mqty(m+i \\ i)}[\eta(t)]^{\frac{n+m}{2}}[1-\eta(t)]^i
\end{equation}
Putting this  $\rho_{n,m}(t)$ into Eq.~\ref{wigfunc} we get the Wigner function. The negativity of the Wigner function in phase space is evident form Fig.~\ref{fig:wigneg}(a) for $\alpha=1m=1,q=0.8,\gamma=0.5,t=0.5$. To see the negativity of the Winger function and the effect of photon loss on the negativity for a range of parameters we use a different approach. We study the behavior of the negativity by calculating the negative volume of the Wigner function \cite{Kenfack2004} in the phase space defined by
\begin{equation}
    \delta(t)=\int\abs{W(\beta,t)}d^2\beta-1
\end{equation}
Though Wigner negative volume is not a linear measure of nonclassicality \cite{Kenfack2004}, it is still a measurable quantity in experiments \cite{Lvovsky2000,Banaszek1999}. Hence it can be used to quantify the effects of the parameters under study.  The result of the Wigner negative volume is shown in Fig.~\ref{fig:wigneg}. We can clearly see from the Fig.~\ref{fig:wigneg}(b) and (c) the negative volume $\delta(t)$ approaches to zero indicating the loss of quantum information (nature) as a result of the decoherence effect. The increasing number of photon addition leads to more $\delta$ initially (at $t=0$) but the decay is also faster for the large $m$ values as seen from Fig.~\ref{fig:wigneg}(b). The variation of $\delta(t)$ is very less due to $q-$deformation for low $m$ numbers. For large values for example $m=5$ the change in  $\delta(t)$ due to the variation of $q$ parameter is noticeable, as shown in  Fig.~\ref{fig:wigneg}(c).The combined influence of varying  $m$ and $q$ can be illustrated through an appropriate choice of parameters. In particular, by fixing 
$\alpha=0.5,\gamma=0.1,m=5$ for different values of $q$, the effect is clearly depicted in Fig.~\ref{fig:wigneg}(d).

\begin{figure*}
  \centering
   (a) \includegraphics[width=0.4\linewidth]{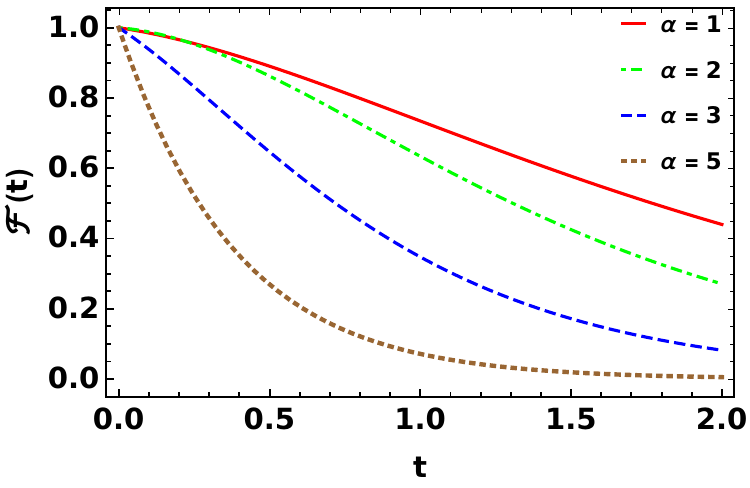}
    (b)\includegraphics[width=0.4\linewidth]{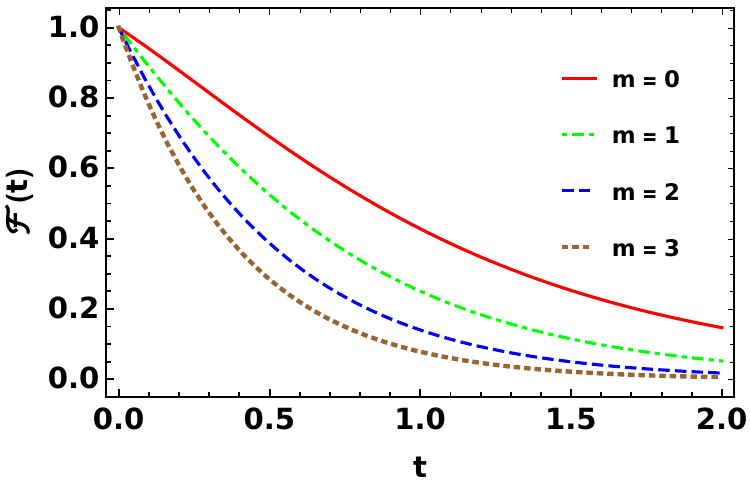}
    (c)\includegraphics[width=0.4\linewidth]{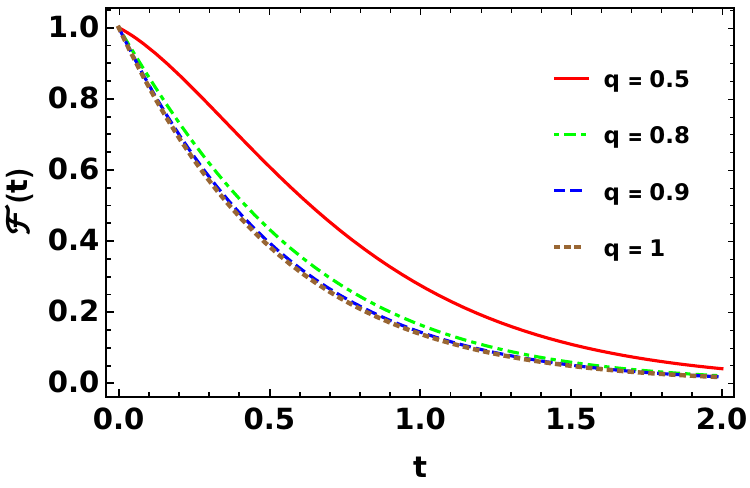}
    (d)\includegraphics[width=0.4\linewidth]{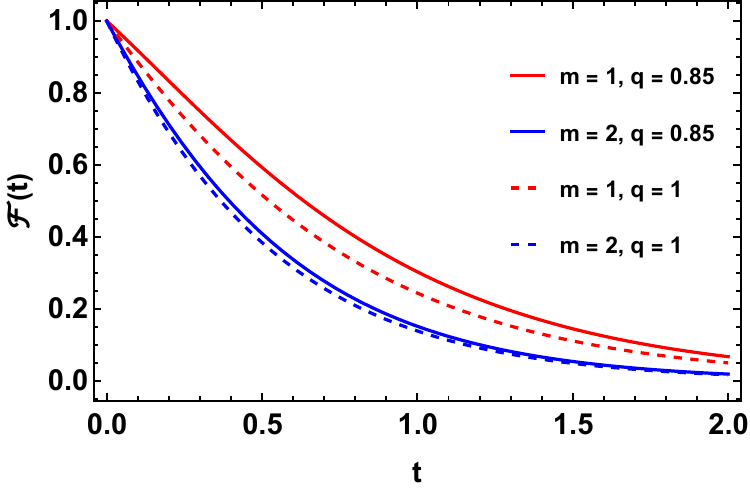}

    \caption{Time evolution ($\gamma=0.5$) of the Fidelity $\mathcal{F}$ of the state DPACS for a range of deformation parameters, considering $q-$deformation oscillator (a) $m=1,q=0.8$ for different values of $\alpha$ (b) $q=0.95,\alpha=3$ for different values of $m$ (c) $m=2,\alpha=3$ for different values of $q$  }
    \label{fid}
\end{figure*}
\subsection{Fidelity}

Fidelity is the measure of overlap between two states. In our context, fidelity quantifies how the state under decoherence $\hat{\rho}(t)$ differs from the initial state $\hat{\rho}(0)$. It is defined as

\begin{equation}
    \mathcal{F}(t)=Tr[\hat{\rho}(t)\hat{\rho}(0)]
\end{equation}
From Eq.~\ref{damp1} and \ref{kir1} the expression for the fidelity is

\begin{equation}
\mathcal{F}(t)=\sum_{j=0}^{\infty}\abs{\sum_{k=j}^{\infty}C^{*}_{k-j}\, C_{k}\sqrt{\mqty(k \\ j)}[\eta(t)]^{\frac{k-j}{2}}[1-\eta(t)]^{\frac{j}{2}}}^{2}
\end{equation}
In the Fig.~\ref{fid} the fidelity of DPACS under the dissipative evolution with the initial state is shown.  For each choice of the parameters, the fidelity eventually decreases to zero. Depending upon the dissipation rate $\gamma$ the time axis is scaled. A comparison between the Figs. \ref{fid}(a)–(c) shows that adjusting the parameter $\alpha$ slows down the decay of fidelity more effectively than modifying $m$ or $q$. When $\alpha$ is large, the coherent state spreads out more in phase space, so even a small amount of damping takes a bigger “bite” out of it. Because of that, the evolved state drifts away from the original one much more quickly, and the fidelity drops faster. For small $\alpha$, the state is tightly packed and closer to the vacuum, so damping hardly changes it and the fidelity stays high for longer. This suggests that the state’s coherence is more resistant when $\alpha$ is varied. In addition, Fig. \ref{fid}(d) shows that states with stronger $q$-deformation tend to retain their coherence for a longer duration, pointing to more robustness to the dissipative environment.

\subsection{Entanglement}
\begin{figure}
    \centering
    (a)\includegraphics[width=0.68\linewidth]{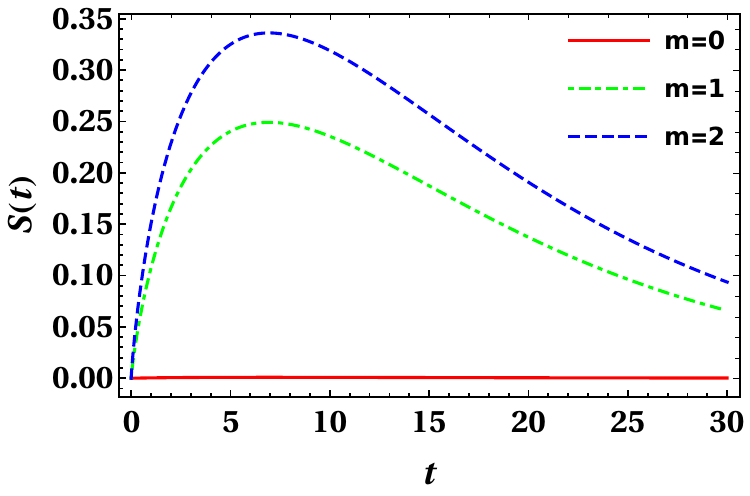}
    (b)\includegraphics[width=0.68\linewidth]{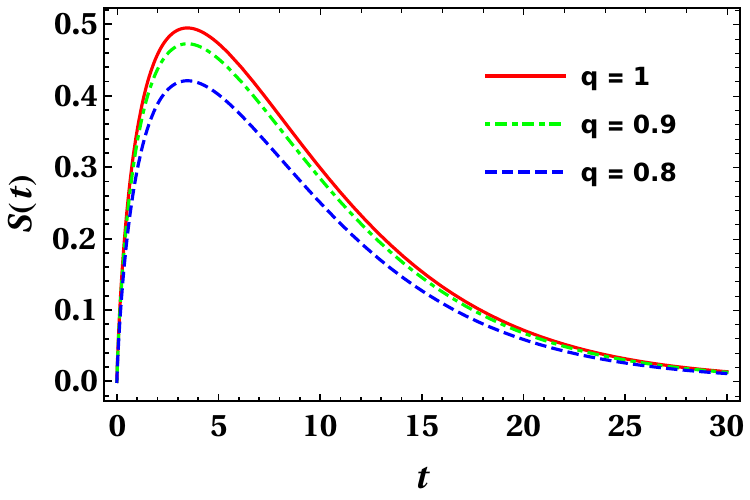}
   
    \caption{Time evolution of the von Neumann entropy $S(t)$ for a range of parameters, considering $q-$deformation oscillator (a) $\alpha=1,q=0.5,\gamma=0.1$ for different values of $m$ (b) $\alpha=1,m=5,\gamma=0.2$ for different values of $q$}
    \label{fig:entrophy}
\end{figure}

Quantum entanglement is another important feature that serves as an independent measure of the nonclassical nature of a quantum state \cite{Kim2002}. Preserving entanglement against decoherence due to environmental interaction remains a major experimental challenge, as entanglement is an important resource for quantum technologies. Thus we investigate the evolution of entropy of our state DPACS under the decoherence and the effect of parameter $m$ and $q$. To quantify the entanglement we use the on Neumann entropy of the density matrix $\hat{\rho}(t)$ as 
\begin{equation}
    S(t)=-Tr[\hat{\rho}(t)ln(\hat{\rho}(t))]=-\sum\lambda_i(t) ln(\lambda_i(t))
\end{equation}
here $\lambda_i$ are the eigenvalues of the reduced density matrix of the system $\hat{\rho}(t)$. The result is shown in Fig.~\ref{fig:entrophy}. As seen from the figures the entropy rises initially and at large time it decays down to zero. Thus decoherence enhances the corelation between the state DPACS and the bath initially then at sufficiently large time; state of the system disentangles itself from the bath becoming pure state. This matches with our previous discussion in Sec.~\ref{PNDD}, specifically in Fig.~\ref{fig:PN(T)}(d-f) that at $t \to \infty$ the $P_{(n=0)}(t) \to 1$ , that means the system leaves all the photon to the bath leaving only the pure state $\ket{0}$. Also the $\ket{0}$ is a classical state with Wigner function as gaussian distribution with Possonian photon statistics ie, $\mathcal{Q}=1$. This is with the agreement of our finding in Sec.~\ref{mandel_q} that at large times $\mathcal{Q}(t)\to1$ for DPACS under decoherence.\\

The q-deformation suppresses higher photon-number components of the photon-added coherent state, resulting in a photon distribution concentrated in the few-photon regime. Consequently, the photon-loss channel generates less statistical mixing during evolution, leading to a reduced von Neumann entropy. This indicates that the deformation enhances the robustness of the nonclassical features of the state against decoherence.


\section{Conclusion}
\label{IV}
In this work we build deformed version of photon added coherent state using $f-$oscillator approach and take the deformation function obeying $q-$oscillator algebra. We investigate the behavior of the state in a dissipative environment, modeled as a photon-loss channel. We subsequently analyzed various nonclassicality parameters, such as the Mandel $\mathcal{Q}$ parameter and the Wigner function , von Neumann entropy, to examine their evolution under the dissipative interaction. We have shown from Mandel $\mathcal{Q}$ parameter that adding more photons leading to more nonclassicality and robustness against decoherence. Simialr effect can be seen due to change in the deformation parameter. We show that the deformation parameter squeezes the photon-number distribution toward lower Fock states, resulting in a reduced von Neumann entropy and thereby indicating enhanced robustness against decoherence.

\section*{ACKNOWLEDGEMENT} A.D. would like to acknowledge UGC JRF, Govt. of India for financial support.


\end{document}